\documentclass[journal]{IEEEtran}
\usepackage{amsmath,amsfonts}
\usepackage{algorithmic}
\usepackage{algorithm}
\usepackage{array}
\usepackage[caption=false,font=normalsize,labelfont=sf,textfont=sf]{subfig}
\usepackage{textcomp}
\usepackage{stfloats}
\usepackage{url}
\usepackage{verbatim}
\usepackage{graphicx}
\usepackage{cite}
\usepackage{amsthm}
\usepackage{wrapfig}

\theoremstyle{definition}

\begin{document}
%
\title{Large Language Model Federated Learning with Blockchain and Unlearning for Cross-Organizational Collaboration}


%
%
%

\author{Xuhan Zuo, Minghao Wang, Tianqing Zhu$^*$, Shui Yu,~\IEEEmembership{Fellow,~IEEE,} Wanlei Zhou,~\IEEEmembership{Fellow,~IEEE}
\thanks{$^*$Tianqing Zhu is the corresponding author with Faculty of Data Science, City University of Macau, Macao (E-mail: tqzhu@cityu.edu.mo)}

\thanks{Xuhan Zuo and Shui Yu are with School of Computer Science, University of Technology Sydney, Ultimo 2007, Australia (E-mail: Xuhan.Zuo-1@student.uts.edu.au; Shui.Yu@uts.edu.au)}
\thanks{Minghao Wang, and Wanlei Zhou are with the Faculty of Data Science, City University of Macau, Macao (E-mail: sydminghao@gmail.com; wlzhou@cityu.edu.mo)}
}

%
%

\markboth{Journal of \LaTeX\ Class Files,~Vol.~14, No.~8, August~2015}%
{Shell \MakeLowercase{\textit{et al.}}: Bare Demo of IEEEtran.cls for IEEE Journals}
%



\maketitle

\begin{abstract}
Large language models (LLMs) have transformed the way computers understand and process human language, but using them effectively across different organizations remains still difficult. When organizations work together to improve LLMs, they face several main challenges. First, organizations hesitate to share their valuable data with others. Second, competition between organizations creates trust problems during collaboration. Third, new privacy laws require organizations to be able to delete specific data when requested, which is especially difficult when multiple organizations are learning from shared data. Traditional federated learning approaches do not address these interconnected challenges, particularly in scenarios where participants cannot fully trust each other or the central aggregator. To overcome these limitations, we propose a hybrid blockchain-based federated learning framework that uniquely combines public and private blockchain architectures with multi-agent reinforcement learning. Our framework enables transparent sharing of model update through the public blockchain while protecting sensitive computations in private chains. Each organization operates as an intelligent agent, using Q-learning to optimize its participation strategy and resource allocation, thus aligning individual incentives with collective goals. Notably, we introduce an efficient unlearning mechanism based on Low-Rank Adaptation (LoRA) that enables selective removal of specific data contributions without compromising the model's overall performance. Through extensive experimentation on real-world datasets, we demonstrate that our framework effectively balances privacy protection, trust establishment, and regulatory compliance while maintaining high model performance. Case studies in healthcare and education sectors validate our approach's practical applicability in sensitive domains where data privacy and trust are paramount.

\end{abstract}

%
\IEEEpeerreviewmaketitle

\section{Introduction}
In recent years, we have witnessed an unprecedented transformation in natural language processing, driven largely by the emergency of Large Language Models (LLMs). These models have revolutionized our ability to process and generate human language \cite{hadi2023large}. However, the scale of data and computational resources required for training these models often exceeds what any single organization can provide, making multi-organizational collaboration not just beneficial, but necessary. Moreover, LLMs needs to consume a huge amount of data, the public data may consumed up while private data from organizations are highly needed. This need for collaboration is particularly evident in specialized domains where valuable data is distributed across multiple organizations. For example, hospitals possess rich repositories of medical narratives and clinical documentation that could significantly enhance medical language models. However, direct data sharing may face substantial regulatory barriers and privacy concerns from organizations. 

While federated learning has emerged as a promising solution for collaborative model training \cite{zhang2023fedrecovery}, allowing organizations to develop models without sharing raw data. As highlighted by \cite{s6}, each organization trains the model locally and shares only model updates with a central server for aggregation into a global model. However, our analysis reveals that the intersection of trust, privacy, and regulatory compliance represents a critical bottleneck in this approach. The issue of trust among participating organizations emerges as a fundamental concern \cite{sanchez2024federatedtrust,MSurvey3}. Traditional federated learning approaches often make optimistic assumptions about participant honesty and aggregator trustworthiness. However, in real-world scenarios, organizations can have competing interests and varying levels of commitment to the collaborative process, which can lead to data poisoning or model manipulation~\cite{MSurvey7}.

Therefore, if we would like to propose a large language model based federated learning, some key challenges should be considered.

\begin{itemize}
    \item 1. Privacy concerns of organizations. Beyond trust concerns, organizations must also navigate the complex landscape of data privacy regulations, particularly regarding data removal requests. The "right to be forgotten" enshrined in GDPR \cite{regulation2018general} presents a unique challenge in collaborative model training. 
    
    \item 2. The trustworthiness and transparency of the collaboration between organizations. Organizations face dual challenges in cross-organizational collaboration: they need to ensure transparent verification of model updates and transactions while protecting sensitive model updates from exposure. A single blockchain architecture cannot address the requirements effectively because public blockchains offer transparency but lack privacy protection, while private blockchains provide privacy but limit cross-organizational verification.

\end{itemize}

For the first challenge, we applied machine unlearning to allow all organizations to withdraw any data they have provided. But we have observed that when multiple organizations contribute to a model's development through numerous iterations, ensuring the complete removal of specific individual's data becomes technically challenging and computationally expensive. Therefore, a key innovation in our framework is the integration of an efficient unlearning mechanism based on the Low-Rank Adaptation (LoRA) technique \cite{hu2021lora}. This mechanism addresses one of the most pressing challenges in collaborative AI development: the ability to selectively remove specific data contributions without compromising the entire model. Our implementation not only ensures regulatory compliance but also significantly reduces the computational cost traditionally associated with model unlearning.

For the second challenge, we propose a novel hybrid blockchain-based federated learning framework designed specifically for LLM development in cross-organizational settings. Our approach uniquely combines public and private blockchain architectures to establish a secure and transparent collaboration environment. The public blockchain serves as an immutable record of model updates and transactions, while private blockchain networks enable organizations to share sensitive model updates within controlled groups~\cite{s2,s3}.


The practical impact of our framework extends beyond theoretical contributions. Through extensive experimentation and real-world case studies in education and healthcare sectors, we demonstrate how our approach effectively balances the competing demands of collaboration efficiency, data privacy, and regulatory compliance. Our results suggest that this framework could serve as a foundation for future large-scale collaborative AI development projects, particularly in sensitive domains where data privacy and trust are paramount concerns.

The main contributions of this paper are:

\begin{itemize}

\item An efficient unlearning mechanism based on LoRA for selective data removal, ensuring compliance with data privacy regulations and maintaining user trust.
\item A novel hybrid blockchain-based federated learning framework with multi-agent interactions and unlearning capabilities for secure, transparent, and efficient collaborative LLM training in cross-organizational settings.
\item Extensive experimental evaluations and case studies demonstrating the effectiveness and practical applicability of our framework in real-world scenarios.
\end{itemize}

\section{Related Work}

\subsection{Federated Learning with LLM}
The integration of federated learning with large language models has emerged as a crucial research direction, particularly as organizations face increasing challenges in accessing high-quality public data while possessing valuable private datasets. This intersection presents unique challenges related to computational resources, communication efficiency, and privacy protection.

Zhao et al \cite{zhao2024llm}. propose PPLR (Privacy-Preserving LLM-based Recommendation), addressing two critical challenges in LLM-based federated recommendation: performance imbalance across clients and high resource demands. Their framework introduces a dynamic balance strategy for parameter aggregation and learning speed adjustment, alongside a flexible storage strategy that selectively manages sensitive model layers. While their approach effectively balances client performance and resource efficiency, it doesn't address the broader challenges of cross-organizational trust and transparent verification that our hybrid blockchain framework provides.

Wu et al \cite{wu2024fedbiot}. introduce FedBiOT, focusing on resource-efficient LLM fine-tuning in federated settings. Their novel bi-level optimization approach involves server-side model compression and client-side adapter fine-tuning, effectively addressing the challenge of limited client resources. While this method significantly reduces resource consumption without compromising performance, it doesn't incorporate the comprehensive privacy protection and data removal capabilities that our LoRA-based unlearning mechanism offers.

Kuang et al \cite{kuang2024federatedscope}. highlight the fundamental challenges in federated LLM fine-tuning, particularly emphasizing the need for frameworks that can optimize resource consumption while meeting diverse information protection demands. Their work identifies critical gaps in existing frameworks regarding communication efficiency, task-specific data preparation, and privacy protection. Our work directly addresses these gaps through a combination of blockchain-based verification and efficient unlearning capabilities.

Ye et al \cite{ye2024openfedllm}. present OpenFedLLM, a comprehensive framework for collaborative LLM training that addresses the impending scarcity of high-quality public data. Their framework encompasses federated instruction tuning, value alignment, and supports diverse domain applications with extensive evaluation metrics. Their results demonstrate the superior performance of federated approaches over local training, particularly in specialized domains. However, their focus primarily remains on performance optimization rather than addressing the complex trust dynamics and privacy concerns in cross-organizational collaboration that our work emphasizes.

In summary, while existing research has made significant progress in addressing various aspects of federated LLM training, our framework uniquely combines blockchain-based trust mechanisms with efficient unlearning capabilities to create a more comprehensive solution for cross-organizational collaboration. This approach not only ensures data privacy and model security but also provides the transparency and verification mechanisms essential for sustainable collaborative AI development.

\subsection{Unlearning with LLM}

The challenge of unlearning specific information from large language models (LLMs) has garnered significant attention, especially as the need to remove sensitive or harmful information becomes increasingly important. Several approaches have been proposed to tackle this issue, each with its strengths and limitations.

Liu et al.\cite{liu2024towards} introduce Selective Knowledge Negation Unlearning (SKU), a novel unlearning framework designed to eliminate harmful knowledge while preserving the utility of LLMs on normal prompts. The SKU framework involves a two-stage process: a harmful knowledge acquisition stage followed by a knowledge negation stage. The study demonstrates that SKU effectively balances the trade-off between unlearning harmful content and maintaining model performance on non-harmful prompts. Compared to this approach, our work extends the idea of selective unlearning by incorporating a more granular control mechanism, allowing for the targeted removal of specific data points with minimal impact on overall model utility.

Chen et al.\cite{chen2023unlearn} propose an effective unlearning framework with an unlearning layer specifically designed for both classification and generation tasks. Their approach focuses on the efficient removal of unwanted knowledge from LLMs, emphasizing the importance of computational efficiency and scalability in the unlearning process. While their method is robust in terms of task versatility, our framework offers a more specialized solution tailored to the unique challenges of LLMs used in federated learning environments, ensuring that unlearning is both precise and minimally disruptive to the model's overall functionality.

Yao et al.\cite{yao2023large} pioneer the concept of large language model unlearning, defining the goal of unlearning in LLMs as the ability to produce non-harmful outputs when faced with harmful prompts. They employ a Gradient Ascent (GA) based method to remove harmful content, though this often results in degraded performance on normal prompts. In contrast, our work introduces a more balanced approach, leveraging the LoRA-based forgetting mechanism to ensure that the removal of harmful information does not compromise the model’s ability to respond accurately to benign queries.

Maini et al.\cite{maini2024tofu} present a new benchmark for evaluating unlearning methods in LLMs, specifically focusing on fictitious unlearning, where the model is tested on its ability to forget contrived or synthetic information. This benchmark provides a useful tool for assessing unlearning efficacy, but it is limited to specific types of data. Our work, however, addresses a broader range of real-world unlearning scenarios, particularly in cross-organizational contexts where different organizations may have varying privacy and security requirements.

Eldan et al.\cite{eldan2023s} introduce an innovative network designed to unlearn copyrighted information embedded within LLMs, highlighting the importance of intellectual property protection in the AI space. Their approach is highly relevant in legal contexts, but our work focuses on a wider application, ensuring that LLMs used in collaborative environments can unlearn a variety of sensitive information while maintaining model performance across diverse tasks.

In summary, while existing research has made significant strides in developing methods for unlearning in LLMs, our work offers a comprehensive and flexible solution that is particularly suited for federated learning scenarios. Our approach not only ensures that sensitive information can be effectively unlearned but also maintains the model’s utility and adaptability in dynamic, cross-organizational environments.

\subsection{Blockchain with LLM}
In recent years, there has been a growing interest in leveraging blockchain technology to address various vulnerabilities and enhance the security of large language models (LLMs). The integration of blockchain with LLMs has emerged as a promising approach to mitigate risks such as data leakage, inference attacks, and other adversarial threats.

Luo et al. \cite{luo2023bc4llm} presents a comprehensive survey on the integration of blockchain with LLMs, exploring how blockchain technology can enhance the trustworthiness of LLMs by ensuring data provenance, integrity, and transparency. Their work categorizes blockchain's role in addressing key vulnerabilities of LLMs, such as prompt injection and data poisoning attacks. While their survey provides a broad overview, our work goes a step further by proposing a hybrid blockchain framework specifically designed for cross-organizational LLM federated learning. This framework not only addresses the security concerns highlighted by Luo et al. but also introduces innovative solutions like the LoRA-based data forgetting mechanism, which enhances data privacy and model adaptability in dynamic environments.

Gong \cite{gong2023dynamic} proposes the concept of Dynamic Large Language Models (DLLMs) on blockchain, which evolve post-training by continuous learning during their usage. This approach leverages the decentralized nature of blockchain to create tamper-resistant datasets that can be audited for accuracy. While Gong's work focuses on the dynamic updating of LLMs, our approach extends the security benefits by integrating both public and private blockchains to balance transparency and privacy, particularly in cross-organizational collaborations where data sensitivity varies.

Lin et al.\cite{lin2024blockchain} introduce a blockchain-based trusted federated offloading framework, which utilizes Chameleon Hash (CH) technology to streamline model updates and reduce computational and consensus costs associated with offloading tasks. This framework ensures the integrity and traceability of model updates while incorporating privacy-preserving results. Compared to their approach, our work further enhances privacy and adaptability through the use of LoRA-based data forgetting mechanisms, which allow for selective data removal without compromising overall model performance.

Mbula et al.\cite{mboma2023assessing} explore the potential of blockchain to provide auditability and traceability in LLMs, particularly in defending against prompt injection attacks. Their work highlights the transparency and immutability of blockchain as critical factors in securing LLM interactions. In contrast, our proposed framework not only incorporates these features but also introduces a multi-agent system that optimizes decision-making processes across organizations, further enhancing the security and efficiency of LLM deployment in federated learning scenarios.

Malhotra et al. \cite{malhotra2024blockchain} propose a blockchain-based proof-of-authenticity framework for explainable AI (XAI), utilizing Ethereum smart contracts to ensure secure and auditable transactions. Their framework emphasizes the importance of transparency and traceability in AI systems. Our work builds on these principles by applying them specifically to the LLM domain, where the hybrid blockchain architecture we propose ensures that both public and private data are securely managed and that all model updates are transparently recorded, providing a robust foundation for cross-organizational collaboration.

In summary, while existing research has explored various aspects of integrating blockchain with LLMs, our work distinguishes itself by offering a comprehensive hybrid blockchain framework designed for federated learning in cross-organizational settings. This framework not only addresses common security concerns but also introduces novel mechanisms like LoRA-based data forgetting, enhancing both the privacy and adaptability of LLMs in real-world applications.



\section{Preliminaries}

\subsection{Large Language Models (LLMs)}
Large Language Models (LLMs) are neural network-based models designed to understand, generate, and manipulate human language. The core of most LLMs is the Transformer architecture, which uses self-attention mechanisms to process sequential data.
The self-attention mechanism can be formulated as:
\begin{equation}
Attention(Q, K, V) = softmax(\frac{QK^T}{\sqrt{d_k}})V
\end{equation}
where Q, K, and V are query, key, and value matrices respectively, and $d_k$ is the dimension of the key vectors.
LLMs are typically trained to minimize the negative log-likelihood of the training data:
\begin{equation}
\mathcal{L}(\theta) = -\sum_{i=1}^{N} \log p(x_i|\theta)
\end{equation}
where $\theta$ represents the model parameters, and $x_i$ are the training examples.
In cross-organizational settings, LLMs can be collaboratively trained and utilized, opening up new possibilities for applications such as personalized education and medical decision support.

\subsection{Federated Learning}
Federated learning is a distributed machine learning paradigm that enables multiple parties to collaboratively train a model without the need for direct data sharing \cite{zhang2023privacyeafl}. In a federated learning setting, each participating entity (i.e., agent) maintains its own local dataset and performs model training locally. The locally trained models are then aggregated to update a global model, which is shared among all agents. This process is repeated iteratively until the global model converges or a desired level of performance is achieved.

Formally, the goal of federated learning is to minimize a global objective function $F(\theta)$, which is defined as the weighted average of the local objective functions $F_i(\theta)$ of $N$ agents:
\begin{equation}
F(\theta) = \sum_{i=1}^{N} w_i F_i(\theta)
\end{equation}

where $\theta$ denotes the model parameters, and $w_i$ represents the weight of the $i$-th agent, which is typically proportional to the size of its local dataset.


\subsection{Blockchain Technology}

Blockchain is a decentralized and distributed ledger technology that enables secure, transparent, and tamper-proof record-keeping \cite{zong2023relac,li2023blockchain}. In our framework, we utilize a permissioned blockchain platform designed for enterprise use, which allows for a more flexible and scalable architecture suitable for cross-organizational collaborations.
The key components of our blockchain system include:
\begin{itemize}
    \item Peers: Nodes that maintain the ledger and execute smart contracts.
    \item Orderers: Nodes responsible for the consensus process and creating new blocks.
    \item Channels: Private subnets of communication between specific network members.
\end{itemize}

The ledger in our blockchain system consists of two distinct components:
\begin{equation}
L = {W, S}
\end{equation}
where $L$ is the ledger, $W$ is the world state (current state of the ledger), and $S$ is the blockchain (transaction log).
Transactions in our blockchain follow a specific lifecycle:
\begin{equation}
T = {Proposal, Endorsement, Ordering, Validation}
\end{equation}
The endorsement policy, which specifies the conditions for transaction validation, can be represented as:
\begin{equation}
E = f(S_1, S_2, ..., S_n)
\end{equation}
where $E$ is the endorsement result, $f$ is the policy function, and $S_1, S_2, ..., S_n$ are the signatures of the endorsing peers.
In our hybrid blockchain architecture, we leverage the channel feature to create private data collections for secure data sharing within organizations:
\begin{equation}
PDC = {data, collection_{definition}, endorsement_{policy}}
\end{equation}
where $PDC$ represents a Private Data Collection.
This architecture allows us to maintain the privacy of sensitive data within organizations while still enabling secure cross-organizational collaboration. The public blockchain serves as a transparent and immutable ledger for recording global model updates, while private channels enable secure sharing of proprietary data and model updates within respective consortia.

\subsection{Multi-Agent Systems and Q-Learning}
Multi-agent systems (MAS) are a subfield of artificial intelligence that focuses on the study of intelligent agents and their interactions in complex environments \cite{wooldridge2009introduction}. In an MAS, each agent is an autonomous entity with its own goals, beliefs, and decision-making capabilities. Agents can interact with each other and with their environment to achieve their objectives, often through cooperation, coordination, and negotiation.

Q-learning is a popular reinforcement learning algorithm that has been widely used in multi-agent settings \cite{watkins1992q}. In multi-agent Q-learning, each agent maintains a Q-table that stores the expected cumulative rewards (Q-values) for taking a particular action in a given state. The Q-values are updated using the following equation:
\begin{equation}
Q(s,a) \leftarrow Q(s,a) + \alpha [r + \gamma \max_{a'} Q(s',a') - Q(s,a)]
\end{equation}

where $s$ and $a$ denote the current state and action, respectively; $s'$ represents the next state; $r$ is the immediate reward; $\alpha$ is the learning rate; and $\gamma$ is the discount factor that balances the importance of immediate and future rewards.

In our proposed framework, we adopt a multi-agent Q-learning approach to model the decision-making processes of participating organizations in the federated learning setting. Each organization is treated as an agent that aims to maximize its own utility while contributing to the collaborative learning process. The Q-learning algorithm enables agents to learn the optimal strategies for participating in the federated learning process, such as determining the amount of resources to contribute and the level of data sharing, based on the rewards they receive.

By incorporating multi-agent Q-learning into our hybrid blockchain-based framework, we can capture the complex dynamics and interactions among participating organizations and design effective incentive mechanisms to encourage honest participation and fair resource contribution. This integration of Q-learning with federated learning and blockchain technology provides a powerful and adaptive approach for enabling secure, transparent, and incentive-aligned collaborative learning with LLMs in cross-organizational settings.


\section{Problem Definition and System Model}

\subsection{Problem Definition}
The integration of Large Language Models (LLMs) with federated learning in cross-organizational collaborations introduces several critical challenges that require novel solutions:
First is the secure cross-organizational collaboration challenge: For N organizations $O = {O_1, O_2, ..., O_N}$, each with private dataset Di, organizations need to collaboratively train an LLM while ensuring Di never leaves Oi. For example, healthcare providers have patient records that could improve medical language understanding, but direct data sharing is restricted by privacy regulations. Traditional federated learning approaches fail to provide sufficient security guarantees when organizations have competing interests or when the central aggregator cannot be fully trusted.

Second is the model update verification challenge: For model updates $U = {U_1, U_2, ..., U_N}$ from $N$ organizations, we need to verify the authenticity and quality of each $U_i$ without accessing the original training data Di. Unlike conventional federated learning settings where participants are typically trusted, our cross-organizational scenario faces potential risks of malicious updates or model tampering. Organizations may intentionally contribute low-quality updates or manipulate the training process for their benefit.

Third is the dynamic data management challenge: When an organization $O_i$ requests to remove a subset of data $D_f \subset D_i$ from the trained model $LLM_g$, the system needs to efficiently transform the model to $LLM_u$ while maintaining performance on the remaining data $D \setminus Df$. This is particularly complex in federated learning settings because the data's influence is distributed across model updates from multiple training iterations. Traditional approaches requiring complete model retraining are impractical in our cross-organizational setting.

Fourth is the organizational autonomy and incentive challenge: Each organization $O_i$ operates with utility function $U_i(r_i, c_i)$, where $r_i$ represents the resources contributed and $c_i$ represents the benefits received from participation. Without proper incentive mechanisms, organizations might adopt strategic behaviors that benefit themselves at the expense of global model performance. The system must ensure that honest participation and high-quality contributions form the dominant strategy for all participants.

These challenges are fundamentally interconnected. For instance, implementing robust verification mechanisms might increase computational overhead, potentially affecting organization's willingness to participate. Similarly, enabling flexible data removal could make it more difficult to maintain model performance and verify contribution quality.

\subsection{System Model}

Our system model combines blockchain technology, multi-agent interactions, and unlearning mechanisms to facilitate secure, efficient cross-organizational collaboration in LLM training. The architecture integrates multiple components to address the challenges outlined in our problem definition.

At its foundation lies the agent architecture, where participating organizations register on the public blockchain and establish their identities through secure $JWT$ tokens. The registration process validates each organization's credentials and assigns unique identifiers, enabling them to participate in the collaborative training process. Organizations with substantial data volumes establish private blockchain environments to enhance training efficiency and protect sensitive information.

The hybrid blockchain structure serves as the backbone of our system, with the public chain maintaining an immutable ledger $L$ comprising world state $W$ and transaction log $S$. This ledger records all model updates and cross-organizational transactions. The private blockchains, defined by $PDC = \{data, collection\_definition, endorsement\_policy\}$, enable organizations to process sensitive data and conduct preliminary computations within protected environments.

Model updates flow through a carefully designed transaction lifecycle $T = \{Proposal, Endorsement, Ordering, Validation\}$. When an organization proposes a model update, it must first receive endorsement from peers according to policy $E = f(S_1, S_2, ..., S_n)$. The endorsed update then undergoes ordering and final validation before integration into the global model. This process ensures the authenticity and quality of all contributions while maintaining transparency.

In the training phase, organizations utilize Q-learning strategies to optimize their participation. The learning process guides decisions about resource allocation and model contribution timing. When the private chain reaches its specified epoch $N_{private\_epoch}$, the locally trained model $LLM_p$ undergoes aggregation within the private chain before secure transmission to the public chain.

Our system incorporates an efficient unlearning mechanism based on LoRA adaptation. When an organization requests data removal, the process transforms the global model $LLM_g$ using carefully tuned parameters $\lambda$ over $E_u$ epochs. This mechanism ensures thorough data removal while preserving model utility on remaining data.

The interaction between these components creates a dynamic environment where organizations can collaboratively enhance the global model while maintaining control over their sensitive data. The public blockchain ensures transparency and accountability, while private chains protect organizational interests and enable efficient local computations.
\section{A Hybrid Blockchain-based Federated Learning Framework}
\subsection{Overview}
Our proposed framework introduces a novel hybrid blockchain architecture that seamlessly integrates public and private blockchains to facilitate secure and efficient cross-organizational collaboration using Large Language Models. The framework ensures transparency, traceability, and data privacy protection while enabling the effective sharing and utilization of data across multiple organizations. The key components of our framework include client registration, global model upload, private blockchain establishment, federated learning training process, private blockchain aggregation, unlearning process using LoRA for forgetting, unlearning verification and submitting unlearning results, and public blockchain aggregation.

In our framework, we introduce a multi-agent system where each participating organization is represented by an agent. These agents are responsible for managing the local training process, contributing to the global model, and interacting with the blockchain network. The agents employ Q-learning, a reinforcement learning technique, to make optimal decisions based on the current state of the system and the rewards received for their actions.

\begin{figure}
\centering
\includegraphics[width=0.5\textwidth]{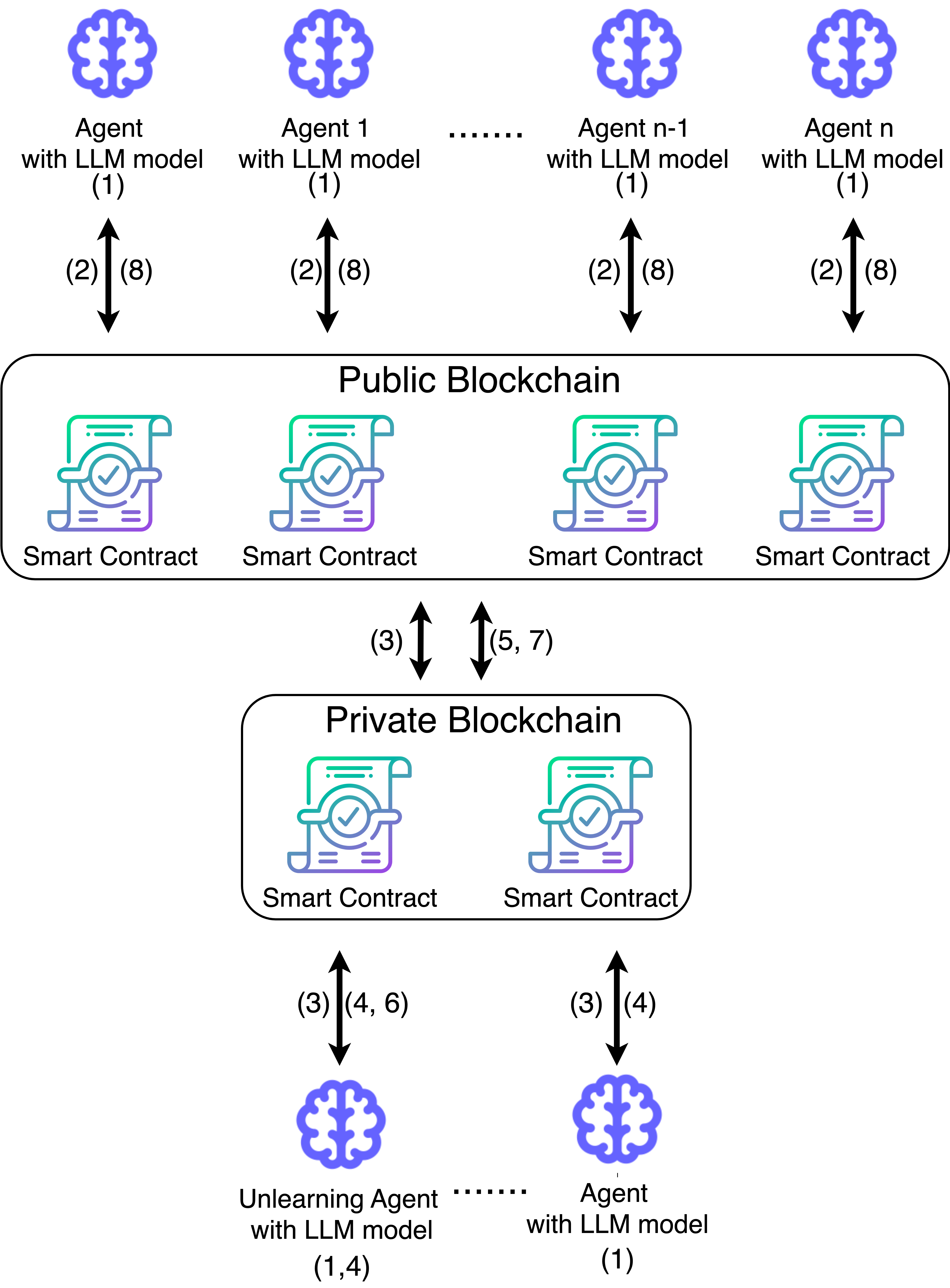}
\caption{Overview and process of our proposed system. (1) Client register. (2) Global model upload. (3) Private blockchain establish. (4) Federated learning training process. (5) Private blockchain aggregation. (6) Unlearning process using LoRA. (7) Unlearning verification and submitting. (8) Public blockchain aggregation. }
\label{public}
\end{figure}

Figure~\ref{public} illustrates the procedure of our proposed federated learning system with public and private blockchains and multi-agent interactions. Initially, all clients must register on the public blockchain. During registration, clients specify their affiliated organization or company. Once registration is complete, the agent uploads the global model to the public chain. Companies with a large number of clients then establish their private chains to enhance the federated learning training process efficiency. The federated learning training epoch is set within the smart contract.

When the private chain's training process reaches the specified epoch, model aggregation commences in the private chain. The private model, $LLM_p$, is then sent to the public chain for further aggregation. If an organization requests to remove their data or model updates, the unlearning process is triggered using LoRA for forgetting. The unlearning results are verified and submitted to the public chain. Upon completion of the model aggregation and unlearning verification processes, the smart contract updates the final global model, $LLM_f$, which the agent can then obtain. The following sections provide a detailed explanation of each component in our proposed framework.

\subsection{Client and Agent Register}
The client and agent registration process is the initial step in our framework. Each participating organization must register as a client to join the collaborative network, while agents are responsible for managing the local training process and interacting with the blockchain network. During registration, both clients and agents provide their unique identifiers and establish secure communication channels using cryptographic techniques such as public-key cryptography. This step ensures that only authorized organizations and agents can participate in the federated learning process and access the shared resources.
\begin{algorithm}
\caption{Client and Agent Register}
\begin{algorithmic}[1]
\REQUIRE $E_{name}$, $Role$, $Org$
\ENSURE $RegSuccess$, $jwt$
\STATE $RegSuccess$ = False;
\STATE Check $Org$;
\IF {$E_{name} \in E_{pool}$}
\STATE \textbf{return} $E_{name}$ already existed.
\ENDIF
\STATE $P_k, S_k \gets keyGen()$;
\STATE $jwt \gets P_k, S_k$;
\STATE $E_{name} \gets jwt$;
\IF {$Role$ is $Client$}
\STATE $Pool_c \gets E_{pool} \cup ID_{ci}$;
\ELSIF {$Role$ is $Agent$}
\STATE $Pool_a \gets E_{pool} \cup ID_{ai}$;
\ENDIF
\STATE $RegSuccess$ = True;
\RETURN  $RegSuccess$, $jwt$
\end{algorithmic}
\end{algorithm}

The algorithm begins by checking if the entity's unique identifier ($E_{name}$) is already present in the entity pool ($E_{pool}$). If the identifier exists, the registration halts, indicating that the entity already exists. The algorithm also verifies the organization ($Org$) associated with the entity.

If the entity is new, the algorithm proceeds to generate a public-secret key pair using the $keyGen()$ function. With these keys, it then creates a JSON Web Token (JWT) for the entity. This JWT, along with the entity ID, is securely stored, effectively registering the entity.

The algorithm then checks the role of the entity. If the role is $Client$, the client pool ($Pool_c$) is updated to include the new client ID. If the role is $Agent$, the agent pool ($Pool_a$) is updated to include the new agent ID.

Finally, the $RegSuccess$ indicator is set to true, and both $RegSuccess$ and the generated $jwt$ are returned, signifying the entity's successful registration and their secure token for future communications.

By incorporating both client and agent registration into this process, the algorithm ensures that all participating entities are properly authenticated and authorized to participate in the federated learning process while maintaining the security and integrity of the system.

\subsection{Global Model Upload}
After successful registration, the agent can upload the global model to the public chain. The global model upload process ensures that all participating organizations have access to the initial model for collaborative training.
\begin{algorithm}
\caption{Global Model Upload}
\begin{algorithmic}[1]
\REQUIRE $jwt$, $LLM_g$
\ENSURE $UploadSuccess$, $LLM$
\STATE $UploadSuccess$ = False;
\IF {$jwt$ is invalid}
\STATE \textbf{return} $jwt$ expired
\ENDIF
\STATE $LLM \gets LLM_g$;
\STATE $UploadSuccess$ = True;
\RETURN  $UploadSuccess$, $LLM$
\end{algorithmic}
\end{algorithm}

The algorithm starts by verifying the validity of the agent's JWT. If the token is invalid or has expired, the process is terminated, and an error message is returned. Upon successful authentication, the global model $LLM_g$ is uploaded to the public chain as $LLM$. The $UploadSuccess$ indicator is set to true, and both $UploadSuccess$ and $LLM$ are returned.

This part of the framework ensures that the global model is securely uploaded to the public chain by an authorized agent, making it accessible to all participating organizations for collaborative training. The process is straightforward and includes the necessary security checks to maintain the integrity of the system.

\subsection{Private Blockchain Establish}

After the global model is successfully uploaded to the public chain, organizations with a large number of clients establish their private blockchains. The private blockchain serves as a secure and tamper-proof ledger for storing and managing the organization's sensitive data and model updates. Ensure data privacy by restricting access to authorized parties within the organization.
\begin{algorithm}
\caption{Private Blockchain Establish}
\begin{algorithmic}[1]
\REQUIRE $jwt$, $LLM_g$
\ENSURE $EstablishSuccess$, $LLM_p$
\STATE $EstablishSuccess$ = False;
\IF {$jwt$ is invalid}
\STATE \textbf{return} $jwt$ expired
\ENDIF
\STATE $LLM_p \gets LLM_g$;
\STATE $EstablishSuccess$ = True;
\RETURN  $EstablishSuccess$, $LLM_p$
\end{algorithmic}
\end{algorithm}

The algorithm verifies the validity of the organization's JWT. If the token is invalid or has expired, the process is terminated, and an error message is returned. Upon successful authentication, the global model $LLM_g$ is uploaded to the private chain as $LLM_p$. The $EstablishSuccess$ indicator is set to true, and both $EstablishSuccess$ and $LLM_p$ are returned.

This part of the framework ensures that organizations with a large number of clients can establish their private blockchains to securely store and manage their sensitive data and model updates. The process includes necessary security checks to maintain the privacy and integrity of the organization's data while allowing them to participate in the federated learning process.

\subsection{Multi-Agent Federated Learning Process on Private Chain}
The multi-agent federated learning process on the private chain enables organizations with a large number of clients to collaboratively train the LLM without directly sharing their sensitive data. Each agent within the organization participates in the training process by leveraging its local data and computational resources. The agents train the model locally and share only the model updates with the organization's private chain. This approach ensures data privacy while benefiting from the collective knowledge of all agents within the organization.
\begin{algorithm}
\caption{Multi-Agent Federated Learning Process on Private Chain}
\begin{algorithmic}[1]
\REQUIRE $Private_{epoch}$, $LLM_p$
\ENSURE $TrainSuccess$
\STATE $TrainSuccess$ = False;
\FOR {$epoch = 1$ to $Private_{epoch}$}
\FOR {each agent $A_i$ in the organization}
\STATE $A_i$ receives $LLM_p$ from the private chain
\STATE $A_i$ trains $LLM_p$ using local data and Q-learning strategy
\STATE $A_i$ sends updated model $LLM_{p,i}$ to the private chain
\ENDFOR
\STATE Aggregate $LLM_{p,i}$ from all agents to update $LLM_p$ on the private chain
\ENDFOR
\STATE $TrainSuccess$ = True;
\RETURN  $TrainSuccess$
\end{algorithmic}
\end{algorithm}

The algorithm begins by setting the $TrainSuccess$ indicator to false. It then iterates for the specified number of $Private_{epoch}$. Within each epoch, the algorithm loops through each agent $A_i$ in the organization. Each agent receives the current $LLM_p$ from the private chain, trains the model using its local data and Q-learning strategy, and sends the updated model $LLM_{p,i}$ back to the private chain. The Q-learning strategy enables agents to make optimal decisions based on the current state of the system and the rewards received for their actions, such as contributing high-quality data or model updates.

After all agents have completed their training for the current epoch, the algorithm aggregates the updated models $LLM_{p,i}$ from all agents to update the $LLM_{p}$ on the private chain. Once all epochs are completed, the $TrainSuccess$ indicator is set to true, and $TrainSuccess$ is returned.

This modified version of the Federated Learning Training Process emphasizes the multi-agent approach, where each agent within the organization contributes to the collaborative learning process using Q-learning strategies. The training occurs on the private chain, allowing organizations with a large number of clients to maintain data privacy and security while benefiting from the collective knowledge of their agents.

\subsection{Private Blockchain Aggregation}
During the federated learning process on the private chain, each agent's model updates are securely stored and aggregated within their respective organization's private blockchain. The private blockchain aggregation mechanism ensures the integrity and traceability of the model updates. It allows each organization to maintain a transparent record of their agents' contributions to the collaborative model while preserving the confidentiality of their sensitive data.
\begin{algorithm}
\caption{Private Blockchain Aggregation}
\begin{algorithmic}[1]
\REQUIRE $jwt$, $Private_{epoch}$, $LLM_p$, $Agg_p$
\ENSURE $AggSuccess$
\STATE $Agg_p$ = False;
\STATE SC realizes that the private chain has reached $Private_{epoch}$;
\IF {$N_p$ reaches $Private_{epoch}$}
\STATE SC aggregates $LLM_p$ from all agents;
\STATE $LLM \gets LLM_p$;
\IF {$jwt$ is invalid}
\STATE \textbf{return} $jwt$ expired
\ENDIF
\ENDIF
\STATE SC sends $LLM$ to $PublicChain$;
\STATE $Agg_p$ = True;
\RETURN  $Agg_p$, $LLM$
\end{algorithmic}
\end{algorithm}

As federated learning progresses on the private chain and the number of epochs $N_p$ reaches the $Private_{epoch}$ setting, the smart contract (SC) aggregates the model updates from all agents on the private chain to obtain $LLM_p$. Additionally, the JWT is verified when the aggregated model is ready to be sent to the public chain. The SC then sends the model to the public chain for further aggregation. The private chain aggregation indicator, $Agg_p$, is set to true, and both $Agg_p$ and the model $LLM$ are returned.

\subsection{Unlearning Process Using LoRA to withdraw data contribution}
The unlearning process enables the selective removal of specific data or model contributions from the federated learning model. When an organization requests to remove their data or model updates, the unlearning process is triggered. We employ Low-Rank Adaptation (LoRA) technology to efficiently forget the specified data without compromising the overall model performance. The unlearning process ensures data privacy and compliance with regulatory requirements.
\begin{algorithm}
\caption{Unlearning Process using LoRA for Forgetting}
\begin{algorithmic}[1]
\REQUIRE $LLM_g$, $D_{forget}$, Learning rate $\eta$, Unlearning epochs $E_u$, LoRA parameters $\lambda$
\ENSURE $params$
\STATE Unlearning Request due to data sensitivity or correction needs;
\STATE Initialize unlearning model $LLM_{local}$ with $LLM_g$;
\STATE Adapter $A$ constructed for $LLM_{local}$ targeting forgetting process;
\FOR {$epoch = 1$ to $E_u$}
\STATE Forward pass with $D_{forget}$ through $LLM_{local}$ to identify features to forget;
\STATE Compute gradients for $LLM_{local}$ emphasizing data points in $D_{forget}$ to be forgotten;
\STATE Apply LoRA to adjust gradients of adapter $A$ using parameters $\lambda$, focusing on unlearning;
\STATE Update $LLM_{local}$'s parameters using the adjusted gradients and learning rate $\eta$, facilitating forgetting;
\ENDFOR
\STATE Calculate the updating $params$ indicative of the forgetting process between $LLM_{local}$ and $LLM_g$;
\RETURN  $params$
\end{algorithmic}
\end{algorithm}

The process begins by initializing a local version of the LLM, denoted as $LLM_{local}$, with the parameters of $LLM_g$. An adapter, $A$, is constructed within $LLM_{local}$ specifically designed to target and facilitate the forgetting of the specified dataset, $D_{forget}$.

The core of the unlearning process involves several training epochs, defined by the parameter $E_u$. In each epoch, a forward pass of $D_{forget}$ through $LLM_{local}$ is performed to identify the characteristics associated with the data points that must be forgotten. Gradients are computed for $LLM_{local}$ with an emphasis on the data to be unlearned. The LoRA technique is applied to the adapter $A$'s gradients using parameters $\lambda$ to focus the unlearning process. With adjusted gradients, $LLM_{local}$'s parameters are updated using the specified learning rate $\eta$. This iterative process gradually leads to the forgetting of the specified data points from $D_{forget}$.

Upon completion of the unlearning epochs, the algorithm calculates the parameters $params$ that indicate the changes made to $LLM_{local}$ compared to $LLM_g$. These parameters represent the outcome of the forgetting process, effectively capturing the essence of what has been unlearned. The algorithm concludes by returning these updated parameters.

\subsection{Unlearning Verification and Submitting Unlearning Results}
The unlearning verification and submission process ensures the integrity and transparency of the unlearning results in the federated learning model. The process involves the agent sending the updated parameters, resulting from the unlearning process, to the smart contract (SC). The SC validates the agent's credentials and evaluates the unlearning results using a validation dataset. If the unlearning results satisfy the verification criteria, the SC submits the updated parameters to the blockchain network.
\begin{algorithm}
\caption{Unlearning Verification and Submitting Unlearning Results}
\begin{algorithmic}[1]
\REQUIRE $params$, Validation dataset $D_{val}$, Agent
\ENSURE $T_{id}$
\STATE Agent sends $params$ to SC;
\IF {Agent's $jwt$ is invalid}
\STATE \textbf{return} Agent identity check failed
\ENDIF
\STATE SC instantiates updated LLM $LLM_{updated}$ with received $params$;
\STATE SC uses $D_{val}$ to evaluate $LLM_{updated}$. Calculates training loss and accuracy to measure unlearning impact.
\IF {Verification criteria are met}
\STATE SC sends $params$ to blockchain network;
\STATE Agents download $params$ from blockchain for weight integration;
\STATE SC ensures updated weights are recorded on blockchain for transparency and traceability;
\STATE SC records Transaction ID $T_{id}$ as proof of submission and integration request;
\ENDIF
\STATE Continue for future federated learning process;
\RETURN $T_{id}$
\end{algorithmic}
\end{algorithm}

The algorithm starts with the agent sending the updated parameters $params$ to the SC. The agent's credentials are validated through their JWT. If the token is invalid, the process halts, indicating a failure in agent identity verification. Upon successful verification, the SC initializes an updated version of the LLM ($LLM_{updated}$) with the new parameters. The SC employs a validation dataset ($D_{val}$) to assess the efficacy of the unlearning process by calculating the loss and precision of training.

If the unlearning results satisfy the predefined verification criteria, the SC submits the updated parameters $params$ to the blockchain network. Agents download these parameters from the blockchain for weight integration into the global model. The SC ensures that the updated weights are recorded on the blockchain, providing transparency and traceability. Additionally, the SC logs a Transaction ID ($T_{id}$), serving as proof of submission and an integration request. The process ends with the return of the transaction ID, indicating the successful verification and submission of the unlearning results.

\subsection{Public Blockchain Aggregation}
The public blockchain aggregation component facilitates the secure and transparent aggregation of the model updates from all participating organizations. The central server collects the model updates from each organization's private blockchain and aggregates them using secure aggregation techniques. The aggregated model updates are then stored on the public blockchain, ensuring transparency and traceability. The public blockchain serves as an immutable record of the collaborative learning process, enhancing trust among the participating organizations.
\begin{algorithm}
\caption{Public Blockchain Aggregation}
\begin{algorithmic}[1]
\REQUIRE $LLM$, $jwt$, $Agg_g$
\ENSURE $epoch$, $Agg_p$
\STATE $Agg_g$ = False;
\IF {$N_g$ reaches $epoch$}
\STATE SC aggregates $LLM$ from all organizations;
\STATE $LLM_g \gets LLM$;
\IF {$jwt$ is invalid}
\STATE \textbf{return} $jwt$ expired
\ENDIF
\ENDIF
\STATE $Agg_g$ = True;
\RETURN $Agg_g$, $LLM_g$
\end{algorithmic}
\end{algorithm}

The algorithm begins by setting the public chain aggregation indicator ($Agg_g$) to false. When the global epoch $N_g$ reaches the previously set epoch, the smart contract (SC) commences model aggregation from all organizations. A JWT must be validated to upload the model. If the token is invalid, an error message is returned. Ultimately, the indicator $Agg_g$ is set to true and returned alongside $LLM_g$.

Following model aggregation on the public chain, a global model update process is initiated, wherein the $LLM_f$ is updated using $LLM_p$ and $LLM_g$.

This completes the detailed explanation of our proposed framework, which leverages a hybrid blockchain architecture to facilitate secure and efficient cross-organizational collaboration using Large Language Models (LLMs) while ensuring data privacy, transparency, and traceability.

\subsection{Case Studies}

To demonstrate the practical applicability and versatility of our proposed framework, we present two case studies that highlight its potential in real-world scenarios. These case studies illustrate how our blockchain-based federated learning framework with unlearning capabilities and multi-agent interactions can be leveraged to address the unique challenges faced by different industries, such as education and healthcare, when collaborating on LLM development.

\subsubsection{Case Study 1: Education University Alliance}
In the first case study, we consider an alliance of universities collaborating to develop an LLM for educational purposes. The LLM aims to assist students, faculty, and researchers by providing personalized learning experiences, intelligent tutoring, and advanced research assistance. Each university possesses a wealth of educational data, including course materials, student interactions, and research publications. However, sharing this data directly among the universities raises concerns about data privacy, intellectual property rights, and the potential misuse of sensitive information.
\paragraph{Implementation of the System}
To realize the education university alliance case study, we adopt our proposed blockchain-based federated learning framework with multi-agent interactions. First, each participating university registers as a client in the system, as shown in Algorithm 1. Then, one university is selected as the agent to upload the initial global LLM model to the public blockchain, as demonstrated in Algorithm 2. Universities with large amounts of data establish their own private blockchains to ensure the privacy and security of their sensitive data (Algorithm 3).

On the private blockchain, each university is represented by an agent that trains the LLM using its local data and Q-learning strategies to make optimal decisions (Algorithm 5). After multiple rounds of training, the aggregated LLM is shared on the private blockchain and then sent to the public blockchain for global aggregation (Algorithm 6).

If a university needs to remove specific data points or model contributions, the unlearning process is triggered (Algorithm 4). The LoRA technique is used to selectively remove data without significantly affecting the overall performance of the LLM. The unlearning results are verified and submitted to the public blockchain (Algorithm 5), ensuring the integrity and transparency of the removal process.

\paragraph{Analysis}
By utilizing our framework, the education university alliance can leverage the collective knowledge and expertise of multiple institutions while preserving data privacy and intellectual property rights. The resulting educational LLM can offer advanced learning experiences and research support to students, faculty, and researchers across the participating universities. The private blockchain ensures that each university's sensitive data remains protected, while the public blockchain facilitates secure collaboration among the different institutions. The multi-agent approach allows each university to make optimal decisions based on its local data and Q-learning strategies, enhancing the overall performance of the LLM. The LoRA-driven unlearning mechanism allows universities to effectively remove specific data as needed while maintaining the overall performance of the LLM.

\paragraph{Challenges and Solutions}
One major challenge in implementing the education university alliance case study is coordinating data sharing and model updates among the different universities. Each university may have varying data formats, privacy requirements, and technical infrastructures. To address this challenge, our framework provides a standardized interface for data sharing and model aggregation, streamlining the collaboration process across different institutions.

Another challenge is ensuring that the unlearning process complies with each university's data retention policies and regulatory requirements. Our framework offers a flexible and verifiable approach to manage diverse data retention needs by using LoRA for fine-grained data removal and verifying the unlearning results on the blockchain.

\paragraph{Discussion}
The education university alliance case study demonstrates the potential application of our blockchain-based federated learning framework with multi-agent interactions in the education domain. By allowing universities to collaborate while protecting data privacy and intellectual property rights, our framework paves the way for developing LLMs for personalized learning, intelligent tutoring, and advanced research. The multi-agent approach enables universities to make optimal decisions based on their local data and learning strategies, while the unlearning mechanism provided by the framework enables universities to manage their data retention policies while ensuring compliance and transparency.

\subsubsection{Case Study 2: Cross-Hospital Collaboration in Healthcare}
The second case study focuses on the collaboration among different hospitals within a healthcare system to develop an LLM for medical decision support and patient care. The LLM aims to assist healthcare professionals by providing evidence-based recommendations, analyzing patient data, and facilitating knowledge sharing among hospitals. Each hospital has a vast repository of medical records, including patient histories, diagnostic images, and treatment outcomes. However, the sensitive nature of this data and the strict regulations governing healthcare information pose significant challenges for collaboration.

\paragraph{Implementation of the System}
To realize the cross-hospital collaboration case study, we adopt a similar approach to the education university alliance case study. Each hospital registers as a client in the system, while one hospital is selected as the agent to upload the initial global LLM to the public blockchain. Hospitals with large amounts of patient data establish their own private blockchains to ensure data privacy and security.

On the private blockchain, each hospital is represented by an agent that trains the LLM using its local patient data and Q-learning strategies to make optimal decisions. The agents collaborate to improve the LLM while maintaining data privacy. The aggregated LLM is shared on the private blockchain and then sent to the public blockchain for global aggregation. If specific patient data or model contributions need to be removed, hospitals can trigger the unlearning process, using LoRA to selectively remove data without affecting the overall performance of the LLM. The unlearning results are verified on the public blockchain, ensuring the integrity of the removal process.

\paragraph{Analysis}
By adopting our framework, the cross-hospital collaboration in healthcare can leverage the collective knowledge and expertise of multiple institutions to develop a powerful medical LLM. The resulting LLM can assist healthcare professionals in making informed decisions, improving patient outcomes, and advancing medical research while maintaining the highest standards of data privacy and regulatory compliance. The private blockchain ensures that each hospital's sensitive patient data remains protected, while the public blockchain facilitates secure collaboration among the hospitals. The multi-agent approach allows hospitals to make optimal decisions based on their local data and Q-learning strategies, enhancing the overall performance of the medical LLM. The LoRA-driven unlearning mechanism allows hospitals to effectively remove specific data as needed while preserving the integrity of the medical LLM.

\paragraph{Challenges and Solutions}
One major challenge in implementing the cross-hospital collaboration case study is ensuring compliance with strict healthcare regulations, such as HIPAA. Our framework addresses this challenge by using private blockchains to isolate sensitive patient data and leveraging secure aggregation protocols to share model updates among hospitals. The blockchain technology also provides an immutable audit trail of data access and sharing activities, ensuring compliance.

Another challenge is managing the different data retention policies and patient consent requirements across hospitals. Our framework offers a flexible and verifiable approach to handle these variations by using LoRA for fine-grained data removal and verifying the unlearning results on the blockchain. This allows hospitals to customize the unlearning process based on their specific data management requirements.

\paragraph{Discussion}
The cross-hospital collaboration case study in healthcare demonstrates the potential application of our blockchain-based federated learning framework with multi-agent interactions in the medical domain. By allowing hospitals to collaborate while maintaining patient privacy and regulatory compliance, our framework paves the way for developing LLMs for medical decision support, patient care, and medical research. The multi-agent approach enables hospitals to make optimal decisions based on their local data and learning strategies, while the unlearning mechanism provided by the framework enables hospitals to manage their data retention policies while ensuring compliance and accountability.
These two case studies showcase the wide-ranging applicability of our blockchain-based federated learning framework with multi-agent interactions in real-world scenarios. By addressing the unique challenges faced by different industries, our framework provides a viable path for the responsible development and deployment of LLMs in critical domains such as healthcare and education.

\section{Privacy and Security Analysis}

\subsection{Privacy Analysis}
The proposed hybrid blockchain-based federated learning framework with multi-agent interactions and unlearning capabilities for Large Language Models (LLMs) is meticulously designed to address the critical privacy challenges associated with collaborative learning in cross-organizational settings. By synergistically integrating the inherent privacy-preserving features of federated learning, the immutability and transparency of blockchain technology, and the efficient data removal mechanisms of unlearning, our approach offers a holistic solution for secure and privacy-centric LLM training.

At its core, federated learning enables the distributed training of LLMs across multiple participants without necessitating the direct exchange of sensitive data \cite{yin2020fedloc}. This decentralized paradigm ensures that each participant maintains control over their proprietary data, significantly mitigating the risks of data breaches and unauthorized access. Mathematically, federated learning can be formulated as an optimization problem that seeks to minimize the global objective function while keeping the data localized:
\begin{equation}
\min_{\theta} \mathcal{L}(\theta) = \sum_{i=1}^{N} \frac{n_i}{n} \mathcal{L}_i(\theta)
\end{equation}

where $\theta$ represents the model parameters, $\mathcal{L}(\theta)$ denotes the global objective function, $\mathcal{L}_i(\theta)$ is the local objective function of the $i$-th participant, $n_i$ is the number of data samples held by the $i$-th participant, $n$ is the total number of data samples across all participants, and $N$ is the total number of participants.

By optimizing the global objective function in this manner, federated learning facilitates the collaborative enhancement of the LLM without exposing raw data, effectively leveraging the distributed data across participants while safeguarding privacy and boosting model performance.

The introduction of multi-agent interactions and Q-learning strategies in our framework further enhances privacy protection by enabling participating organizations to make intelligent decisions about data sharing and model contributions based on their individual privacy constraints and objectives. Each agent autonomously learns the optimal strategies for engaging in the federated learning process, considering factors such as data sensitivity, regulatory compliance, and the potential benefits and risks of collaboration. This decentralized decision-making approach empowers organizations to maintain granular control over their data and reduces the reliance on centralized control mechanisms that may introduce additional privacy vulnerabilities.

Moreover, the integration of blockchain technology in our framework provides an immutable and transparent ledger of all interactions and transactions within the federated learning process, ensuring the integrity and accountability of the collaborative learning process. The utilization of smart contracts automates the enforcement of predefined privacy policies and conditions, guaranteeing that all participants adhere to mutually agreed-upon privacy standards. This automation minimizes the potential for human error and mitigates the risk of unauthorized data access or manipulation.

The unlearning mechanism embedded within our framework empowers participants to selectively remove specific data points or model updates, granting them fine-grained control over their data lifecycle and facilitating compliance with evolving privacy regulations. Theoretically, the unlearning process can be modeled as a constrained optimization problem, wherein the objective is to minimize the impact of the removed data on the model's performance while satisfying the unlearning constraints:
\begin{equation}
\min_{\theta} \mathcal{L}(\theta) = \sum_{i=1}^{N} \frac{n_i}{n} \mathcal{L}_i(\theta) \quad \text{s.t.} \quad \theta \in \Theta_u
\end{equation}
where $\Theta_u$ represents the feasible set of model parameters after unlearning. The goal is to identify the optimal model parameters that minimize the impact of the removed data on the model's performance while adhering to the unlearning constraints. By incorporating this unlearning mechanism, our framework provides participants with a powerful tool to manage their data lifecycle and maintain model performance.

\subsection{Security Analysis}
The integration of blockchain technology and multi-agent interactions in our federated learning framework significantly enhances the security of the collaborative learning process. The immutable nature of blockchain ensures that all model updates and transactions are tamper-proof and easily verifiable, providing a robust defense against malicious actors attempting to manipulate the learning process \cite{politou2019blockchain}.

From a theoretical perspective, the security of a blockchain network can be analyzed through the lens of game theory and consensus mechanisms. In a proof-of-stake (PoS) based blockchain, network security is maintained by requiring participants to stake a portion of their assets as collateral \cite{akbar2021distributed}. This staking mechanism incentivizes participants to act honestly, as any malicious behavior would result in the loss of their staked assets. The security of the network can be modeled as a game between honest and malicious participants, where honest participants aim to maximize their rewards by following the protocol, while malicious participants seek to maximize their gains by deviating from the protocol. The Nash equilibrium of this game represents a state in which no participant can benefit by unilaterally altering their strategy, ensuring the stability and security of the blockchain network.

The multi-agent approach introduced in our framework adds an extra layer of security by enabling participating organizations to independently assess the credibility and trustworthiness of other agents based on their past behavior and contributions. Agents can learn to identify and isolate malicious or free-riding participants, minimizing their impact on the collaborative learning process. This decentralized trust mechanism complements the security features of the blockchain, creating a more resilient and adaptive system that can effectively respond to evolving security threats.

Furthermore, the Q-learning strategies employed by the agents allow them to dynamically adapt their behavior based on the observed security state of the system. Agents can learn to take proactive measures, such as increasing the frequency of model validations or adjusting the staking requirements, to maintain the integrity of the federated learning process in the face of potential attacks. This adaptive security approach enables the system to remain robust and responsive even in the presence of sophisticated adversaries.

Our framework also leverages advanced cryptographic primitives, such as threshold signatures and zero-knowledge proofs, to ensure the integrity and confidentiality of all transactions. Threshold signatures allow for the distributed generation and verification of signatures, eliminating single points of failure and enhancing the resilience of the system against attacks. Zero-knowledge proofs enable participants to validate the correctness of computations without revealing the underlying data \cite{wan2022zk}, preserving privacy while maintaining trust in the federated learning process. By combining these cryptographic techniques with the security features of blockchain and multi-agent interactions, our framework establishes a secure and trustworthy environment for collaborative LLM development.

The utilization of smart contracts further strengthens the security of the system by automating the execution of predefined rules and conditions, minimizing the potential for unauthorized access or manipulation. In our framework, smart contracts govern the federated learning process, enforcing participant adherence to agreed-upon security protocols and facilitating the secure aggregation of model updates. This automated enforcement reduces the risk of human error and malicious behavior, bolstering the overall security of the system.

The decentralized architecture of our framework, enabled by the hybrid blockchain design, eliminates single points of failure and distributes risk across multiple nodes. This distributed approach significantly increases the difficulty for attackers to compromise the entire system, as they would need to control a substantial portion of the participating nodes simultaneously. The probability of a successful attack decreases exponentially with the number of honest nodes in the network, making it practically infeasible in a large-scale, cross-organizational federated learning setting.

In conclusion, our hybrid blockchain-based federated learning framework with multi-agent interactions and unlearning capabilities offers a comprehensive solution for addressing security concerns in cross-organizational LLM training. By harnessing the inherent security features of blockchain technology, multi-agent interactions, and advanced cryptographic techniques, our approach creates a resilient and secure environment for collaborative LLM development. The adaptive security measures enabled by Q-learning strategies and the decentralized trust mechanism further fortify the system's defenses against evolving security threats, ensuring the integrity and reliability of the federated learning process in complex, multi-stakeholder settings.

\section{Performance Evaluation}


\subsection{Experimental Setup}

\textbf{Datasets: } For our experiments, we utilized two distinct datasets: the IMDB dataset for sentiment analysis and a dataset of tweets from Twitter. These datasets were selected based on several important criteria:

\begin{itemize}
    \item \textbf{Relevance to LLM Applications:} The IMDB dataset is a standard benchmark for sentiment analysis, making it ideal for evaluating the performance of our framework in a well-known context. The Twitter dataset, on the other hand, provides real-world text data applicable to various NLP tasks such as sentiment analysis, topic modeling, and text classification.
    \item \textbf{Size and Complexity:} The IMDB dataset consists of 50,000 movie reviews, offering a substantial but manageable volume for federated learning experiments. The Twitter dataset includes a large number of tweets, which helps assess the scalability and efficiency of our framework.
    \item \textbf{Diversity of Content:} The IMDB dataset contains structured, domain-specific reviews, while the Twitter dataset includes a wide range of topics, opinions, and writing styles. This diversity allows us to test the robustness and adaptability of our framework.
    \item \textbf{Sensitive Information:} Tweets often contain personal or sensitive information that users might wish to erase. This makes the Twitter dataset particularly suitable for evaluating our unlearning mechanism's effectiveness in removing specific data points while preserving overall model performance.
\end{itemize}

\textbf{Evaluation Metrics:} Accuracy was chosen as the primary metric for evaluating our experiments. It provides a straightforward measure of the proportion of correctly classified reviews post-unlearning. By comparing accuracy before and after unlearning, we can gauge the framework's effectiveness in forgetting specific data points while retaining overall model performance.

\textbf{Experimental Comparisons:} Given the unique combination of federated learning, blockchain technology, and unlearning capabilities in our approach, there are no direct counterparts in the current literature for comparison. Our framework is pioneering in addressing selective unlearning within a federated learning setup for LLMs, enhanced with blockchain to ensure privacy and security.

For comprehensive evaluation, we varied LoRA hyperparameters such as rank and scaling factor to identify the configurations that best balance unlearning effectiveness and model accuracy. The multi-agent system configuration in our experiments involved each agent (representing an organization) training locally on its data and collaboratively updating the global model via blockchain to ensure transparency and accountability.

\textbf{Hardware and Software Environment:} The experiments were run on a system with an Intel Xeon 6238R processor, 64GB RAM, and an NVIDIA A6000 GPU, using software environments like Ubuntu 20.04, Visual Studio Code, Hyperledger Fabric, FATE, and machine learning frameworks such as PyTorch and TensorFlow.

\subsection{Results and Discussion}

\textbf{IMDB Dataset Results:} We experimented with various LoRA settings on the IMDB dataset. The Retrain from Scratch method served as a benchmark to evaluate our unlearning approach. Table \ref{imdb_results} presents the initial and final accuracies for different LoRA configurations.

\begin{table}[h!]
\centering
\caption{Results on IMDB Dataset}
\begin{tabular}{|c|c|c|}
\hline
\textbf{LoRA Config} & \textbf{Initial Accuracy} & \textbf{Final Accuracy} \\
\hline
r=32, alpha=2, dropout=0.1 & 97.10\% & 0.95\% \\
r=16, alpha=8, dropout=0.1 & 95.95\% & 1.00\% \\
r=16, alpha=4, dropout=0.5 & 98.00\% & 1.10\% \\
r=16, alpha=4, dropout=0.2 & 98.70\% & 1.20\% \\
r=2, alpha=16, dropout=0.1 & 84.41\% & 1.25\% \\
\hline
\end{tabular}
\label{imdb_results}
\end{table}

\textbf{Twitter Dataset Results:} Similarly, we conducted experiments on the Twitter dataset to assess our method's performance. Table \ref{twitter_results} shows the results of these experiments.

\begin{table}[h!]
\centering
\caption{Results on Twitter Dataset}
\begin{tabular}{|c|c|c|}
\hline
\textbf{LoRA Config} & \textbf{Initial Accuracy} & \textbf{Final Accuracy} \\
\hline
r=16, alpha=8, dropout=0.2 & 83.98\% & 7.93\% \\
r=16, alpha=1, dropout=0.1 & 84.68\% & 8.33\% \\
r=2, alpha=1, dropout=0.3 & 74.42\% & 8.41\% \\
r=16, alpha=2, dropout=0.4 & 81.33\% & 8.42\% \\
r=8, alpha=1, dropout=0.1 & 91.00\% & 9.04\% \\
\hline
\end{tabular}
\label{twitter_results}
\end{table}

\subsubsection{Impact of Rank on Unlearning}
\begin{figure*}[htbp]
 \centering
  \begin{minipage}[b]{0.32\textwidth}
    \centering
    \includegraphics[width=\textwidth]{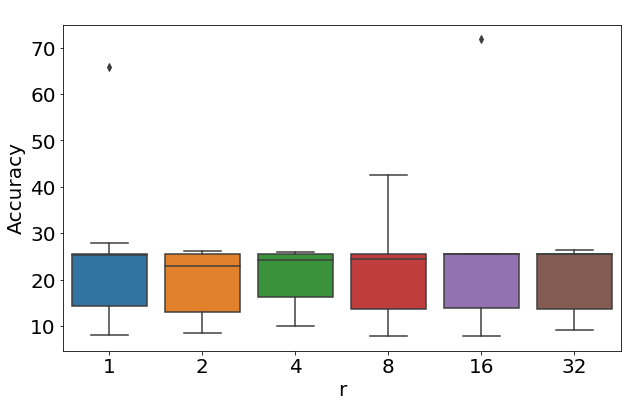}
    \caption{Impact of Different $r$ Values on Accuracy (Twitter)}
    \label{fig:twitter_rank}
  \end{minipage}
  \begin{minipage}[b]{0.32\textwidth}
    \centering
    \includegraphics[width=\textwidth]{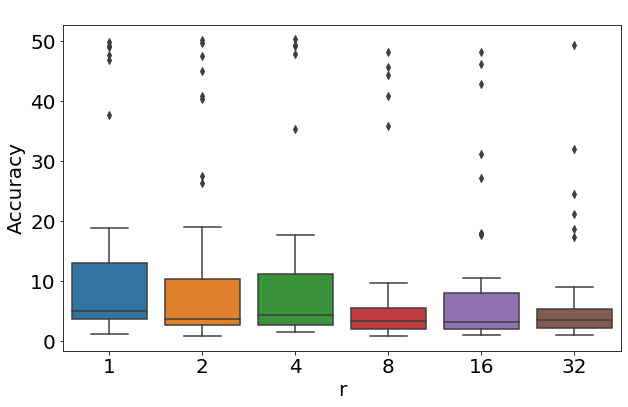}
    \caption{Impact of Different $r$ Values on Accuracy (IMDB)}
    \label{fig:imdb_rank}
  \end{minipage}
  \begin{minipage}[b]{0.32\textwidth}
    \centering
    \includegraphics[width=\textwidth]{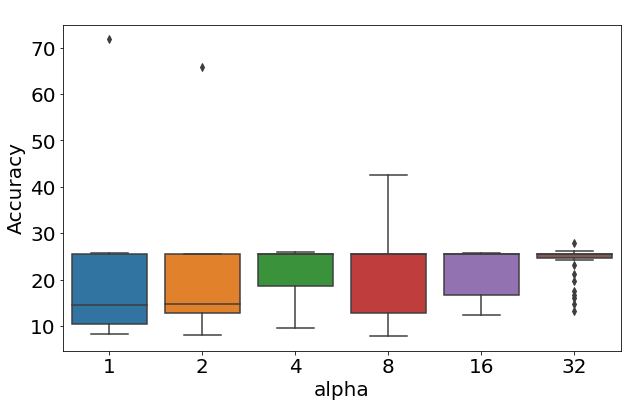}
    \caption{Impact of Different Alpha Values on Accuracy (Twitter)}
    \label{fig:twitter_alpha}
  \end{minipage}

\end{figure*}

Figures \ref{fig:twitter_rank} and \ref{fig:imdb_rank} illustrate how different ranks ($r$) affect accuracy reduction for the Twitter and IMDB datasets, respectively. Higher ranks, such as $r=16$ and $r=32$, generally result in more significant accuracy reductions post-unlearning, indicating better unlearning performance.

\subsubsection{Impact of Alpha on Unlearning}
Figures \ref{fig:twitter_alpha} and \ref{fig:imdb_alpha} show the effects of different alpha values on accuracy reduction for the Twitter and IMDB datasets. Lower alpha values generally lead to more effective unlearning, as evidenced by greater accuracy reductions.



\subsubsection{Impact of Dropout on Unlearning}
\begin{figure*}[htbp]
 \centering
  \begin{minipage}[b]{0.32\textwidth}
    \centering
    \includegraphics[width=\textwidth]{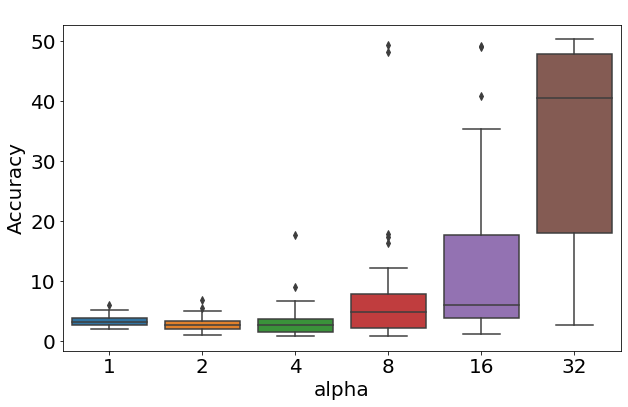}
    \caption{Impact of Different Alpha Values on Accuracy (IMDB)}
    \label{fig:imdb_alpha}
  \end{minipage}
  \begin{minipage}[b]{0.32\textwidth}
    \centering
    \includegraphics[width=\textwidth]{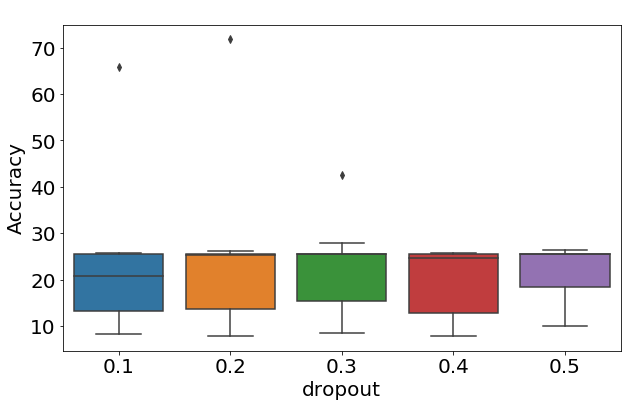}
    \caption{Impact of Different Dropout Values on Accuracy (Twitter)}
    \label{fig:twitter_dropout}
  \end{minipage}
  \begin{minipage}[b]{0.32\textwidth}
    \centering
    \includegraphics[width=\textwidth]{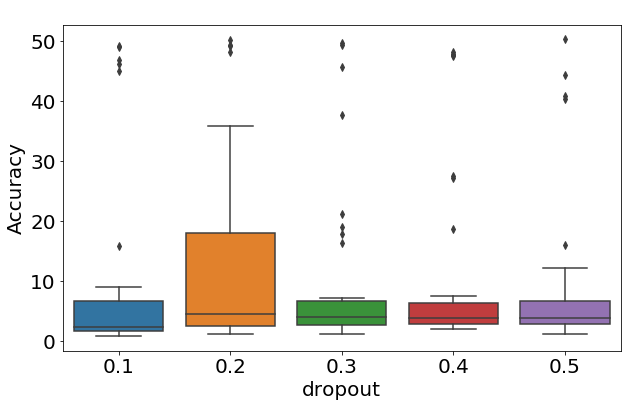}
    \caption{Impact of Different Dropout Values on Accuracy (IMDB)}
    \label{fig:imdb_dropout}
  \end{minipage}

\end{figure*}

Figures \ref{fig:twitter_dropout} and \ref{fig:imdb_dropout} depict how different dropout values influence accuracy reduction. Higher dropout rates ($0.4$ and $0.5$) tend to enhance unlearning performance by introducing more noise during training, thus facilitating better forgetting of specific data.



\subsubsection{Performance Influencing Factors}
Analyzing the impact of alpha, dropout, and $r$ values reveals important insights into the performance of our unlearning method. Both datasets show that lower alpha values and higher dropout rates contribute significantly to improved unlearning by reducing the model's capacity to retain information.

Moreover, higher $r$ values allow the model to capture more diverse information, facilitating better unlearning. These findings underscore the need to carefully tune hyperparameters in our LoRA-based unlearning approach to maximize effectiveness.

\subsubsection{Comparison with Retrain from Scratch}
Table \ref{tab:final_accuracy_comparison} compares our method with the Retrain from Scratch technique for both datasets. Our approach achieves similar final accuracies, demonstrating effective unlearning while being computationally more efficient.

\begin{table}[h!]
\centering
\caption{Final Accuracy Comparison}
\begin{tabular}{|c|c|c|}
\hline
\textbf{Method} & \textbf{Initial Accuracy} & \textbf{Final Accuracy} \\
\hline
Twitter \& Our Method & 83.98\% & 7.93\% \\
Twitter \& Retrain from Scratch & 87.53\% & 7.84\% \\
IMDB \& Our Method & 97.10\% & 0.95\% \\
IMDB \& Retrain from Scratch & 95.60\% & 0.85\% \\
\hline
\end{tabular}
\label{tab:final_accuracy_comparison}
\end{table}

\subsubsection{Blockchain Integration and Performance}
Our study also evaluates the performance implications of integrating a hybrid blockchain structure, combining public and private blockchains, into our federated learning framework with unlearning capabilities. We focused on scalability, transaction throughput, and latency.

\begin{itemize}
\item \textbf{Setup Time:} Initial setup of the hybrid blockchain network took approximately $48$ seconds. This is slightly higher than using only a public blockchain but acceptable considering the long-term benefits in security and privacy.
\item \textbf{Consensus Overhead:} The consensus process added around $6$ seconds due to the additional coordination required between public and private blockchains. This increase is manageable within the federated learning context.
\item \textbf{Transaction Processing:} Average transaction processing time, including model updates and data sharing, was $4$ seconds, demonstrating the hybrid blockchain's efficiency.
\item \textbf{Per-Epoch Duration:} Training duration per epoch remained consistent at $30$-$32$ seconds, even with additional unlearning activities and blockchain coordination, highlighting the system's robustness.
\end{itemize}

Table \ref{tab:time_cost_comparison} compares time costs across different federated learning cycles, showing that while the hybrid blockchain method has a slightly higher initial time cost, it normalizes over iterations, indicating scalability.

\begin{table}[h]
\centering
\caption{Time Cost Analysis for LLM Federated Learning with and without Blockchain Integration}
\label{tab:time_cost_comparison}
\begin{tabular}{|p{3.2cm}|c|c|c|c|}
\hline
\textbf{Method} & $\mathbf{t = 0}$ & $\mathbf{t = 9}$ & $\mathbf{t = 99}$ & $\mathbf{t = 999}$ \\
\hline
Normal Federated Learning for LLMs & $30$s & $300$s & $3000$s & $30000$s \\
\hline
Public Blockchain-Enhanced Method for LLMs & $79$s & $367$s & $3277$s & $32277$s \\
\hline
Hybrid Blockchain-Enhanced Method for LLMs & $84$s & $378$s & $3384$s & $33384$s \\
\hline
\end{tabular}
\end{table}

The hybrid blockchain architecture provides additional data privacy benefits by using private blockchains for sensitive information sharing while maintaining transparency on the public blockchain. This approach balances the need for transparency and accountability with privacy requirements.

Overall, our results demonstrate that integrating a hybrid blockchain into our federated learning framework introduces minimal overhead while ensuring robust and scalable performance. This makes our system a promising solution for secure, transparent, and privacy-preserving federated learning with unlearning capabilities for LLMs.

\subsubsection{Discussion}
Our experiments on the IMDB and Twitter datasets confirm that our method achieves performance levels comparable to the Retrain from Scratch technique in terms of accuracy reduction. The success of our LoRA-based unlearning approach is due to carefully selected and tuned parameters and specific implementation techniques. Our method offers a computationally viable alternative to retraining from scratch, which can be resource-intensive and time-consuming.

Additionally, we evaluated the impact of incorporating blockchain technology into our federated learning framework. The results show that the added blockchain components introduce negligible overhead in terms of setup, consensus, transaction processing, and per-epoch time costs. The system's consistent performance, despite additional unlearning activities, highlights its resilience and scalability.

By balancing performance, privacy, and computational efficiency, our multi-agent blockchain-integrated federated learning framework with unlearning capabilities presents a robust solution for secure and effective LLM training in diverse applications.

\section{Conclusion}
This paper introduces an innovative hybrid blockchain-based multi-agent federated learning framework for training Large Language Models (LLMs) in cross-organizational collaborations, with data unlearning capabilities. Our framework leverages the strengths of both public and private blockchains to create a secure, transparent, and efficient collaborative environment while incorporating multi-agent interactions and efficient data unlearning mechanisms.
Through extensive experiments on IMDB and Twitter datasets, we demonstrate the superior performance of our framework in terms of data privacy protection, collaboration efficiency improvement, and targeted data forgetting. The carefully tuned LoRA hyperparameters enable our approach to efficiently remove target data while maintaining the model's performance on the remaining data. The multi-agent system enhances collaboration efficiency through interactions and knowledge sharing among agents. Furthermore, the hybrid blockchain architecture introduces minimal computational overhead and time cost, highlighting the scalability and robustness of our system.
Compared to existing methods, our framework exhibits significant advantages in terms of computational efficiency, versatility, and adaptability. It provides a secure, transparent, and efficient solution for federated learning of LLMs in cross-organizational settings. Our framework has the potential to drive innovative applications, particularly in scenarios where data privacy and selective data forgetting are of paramount importance.


%





\bibliographystyle{IEEEtran}
\bibliography{IEEEexample}

\begin{thebibliography}{10}
\providecommand{\url}[1]{#1}
\csname url@samestyle\endcsname
\providecommand{\newblock}{\relax}
\providecommand{\bibinfo}[2]{#2}
\providecommand{\BIBentrySTDinterwordspacing}{\spaceskip=0pt\relax}
\providecommand{\BIBentryALTinterwordstretchfactor}{4}
\providecommand{\BIBentryALTinterwordspacing}{\spaceskip=\fontdimen2\font plus
\BIBentryALTinterwordstretchfactor\fontdimen3\font minus \fontdimen4\font\relax}
\providecommand{\BIBforeignlanguage}[2]{{%
\expandafter\ifx\csname l@#1\endcsname\relax
\typeout{** WARNING: IEEEtran.bst: No hyphenation pattern has been}%
\typeout{** loaded for the language `#1'. Using the pattern for}%
\typeout{** the default language instead.}%
\else
\language=\csname l@#1\endcsname
\fi
#2}}
\providecommand{\BIBdecl}{\relax}
\BIBdecl

\bibitem{hadi2023large}
M.~U. Hadi, R.~Qureshi, A.~Shah, M.~Irfan, A.~Zafar, M.~B. Shaikh, N.~Akhtar, J.~Wu, S.~Mirjalili \emph{et~al.}, ``Large language models: a comprehensive survey of its applications, challenges, limitations, and future prospects,'' \emph{Authorea Preprints}, 2023.

\bibitem{zhang2023fedrecovery}
L.~Zhang, T.~Zhu, H.~Zhang, P.~Xiong, and W.~Zhou, ``Fedrecovery: Differentially private machine unlearning for federated learning frameworks,'' \emph{IEEE Transactions on Information Forensics and Security}, 2023.

\bibitem{s6}
\BIBentryALTinterwordspacing
Y.~Zhao, Y.~Qu, Y.~Xiang, M.~P. Uddin, D.~Peng, and L.~Gao, ``A comprehensive survey on edge data integrity verification: Fundamentals and future trends,'' \emph{ACM Comput. Surv.}, vol.~57, no.~1, Oct. 2024. [Online]. Available: \url{https://doi.org/10.1145/3680277}
\BIBentrySTDinterwordspacing

\bibitem{sanchez2024federatedtrust}
P.~M.~S. S{\'a}nchez, A.~H. Celdr{\'a}n, N.~Xie, G.~Bovet, G.~M. P{\'e}rez, and B.~Stiller, ``Federatedtrust: A solution for trustworthy federated learning,'' \emph{Future Generation Computer Systems}, vol. 152, pp. 83--98, 2024.

\bibitem{MSurvey3}
J.~Zhang, L.~Pan, Q.-L. Han, C.~Chen, S.~Wen, and Y.~Xiang, ``Deep learning based attack detection for cyber-physical system cybersecurity: A survey,'' \emph{IEEE/CAA Journal of Automatica Sinica}, vol.~9, no.~3, pp. 377--391, 2021.

\bibitem{MSurvey7}
W.~Zhou, X.~Zhu, Q.-L. Han, L.~Li, X.~Chen, S.~Wen, and Y.~Xiang, ``The security of using large language models - a survey with emphasis on chatgpt,'' \emph{IEEE/CAA Journal of Automatica Sinica}, 2025.

\bibitem{regulation2018general}
G.~D.~P. Regulation, ``General data protection regulation (gdpr),'' \emph{Intersoft Consulting, Accessed in October}, vol.~24, no.~1, 2018.

\bibitem{hu2021lora}
E.~J. Hu, Y.~Shen, P.~Wallis, Z.~Allen-Zhu, Y.~Li, S.~Wang, L.~Wang, and W.~Chen, ``Lora: Low-rank adaptation of large language models,'' \emph{arXiv preprint arXiv:2106.09685}, 2021.

\bibitem{s2}
\BIBentryALTinterwordspacing
Y.~Qu, M.~P. Uddin, C.~Gan, Y.~Xiang, L.~Gao, and J.~Yearwood, ``Blockchain-enabled federated learning: A survey,'' \emph{ACM Comput. Surv.}, vol.~55, no.~4, Nov. 2022. [Online]. Available: \url{https://doi.org/10.1145/3524104}
\BIBentrySTDinterwordspacing

\bibitem{s3}
Y.~Qu, L.~Gao, Y.~Xiang, S.~Shen, and S.~Yu, ``Fedtwin: Blockchain-enabled adaptive asynchronous federated learning for digital twin networks,'' \emph{IEEE Network}, vol.~36, no.~6, pp. 183--190, 2022.

\bibitem{zhao2024llm}
J.~Zhao, W.~Wang, C.~Xu, Z.~Ren, S.-K. Ng, and T.-S. Chua, ``Llm-based federated recommendation,'' \emph{arXiv preprint arXiv:2402.09959}, 2024.

\bibitem{wu2024fedbiot}
F.~Wu, Z.~Li, Y.~Li, B.~Ding, and J.~Gao, ``Fedbiot: Llm local fine-tuning in federated learning without full model,'' in \emph{Proceedings of the 30th ACM SIGKDD Conference on Knowledge Discovery and Data Mining}, 2024, pp. 3345--3355.

\bibitem{kuang2024federatedscope}
W.~Kuang, B.~Qian, Z.~Li, D.~Chen, D.~Gao, X.~Pan, Y.~Xie, Y.~Li, B.~Ding, and J.~Zhou, ``Federatedscope-llm: A comprehensive package for fine-tuning large language models in federated learning,'' in \emph{Proceedings of the 30th ACM SIGKDD Conference on Knowledge Discovery and Data Mining}, 2024, pp. 5260--5271.

\bibitem{ye2024openfedllm}
R.~Ye, W.~Wang, J.~Chai, D.~Li, Z.~Li, Y.~Xu, Y.~Du, Y.~Wang, and S.~Chen, ``Openfedllm: Training large language models on decentralized private data via federated learning,'' in \emph{Proceedings of the 30th ACM SIGKDD Conference on Knowledge Discovery and Data Mining}, 2024, pp. 6137--6147.

\bibitem{liu2024towards}
Z.~Liu, G.~Dou, Z.~Tan, Y.~Tian, and M.~Jiang, ``Towards safer large language models through machine unlearning,'' \emph{arXiv preprint arXiv:2402.10058}, 2024.

\bibitem{chen2023unlearn}
J.~Chen and D.~Yang, ``Unlearn what you want to forget: Efficient unlearning for llms,'' \emph{arXiv preprint arXiv:2310.20150}, 2023.

\bibitem{yao2023large}
Y.~Yao, X.~Xu, and Y.~Liu, ``Large language model unlearning,'' \emph{arXiv preprint arXiv:2310.10683}, 2023.

\bibitem{maini2024tofu}
P.~Maini, Z.~Feng, A.~Schwarzschild, Z.~C. Lipton, and J.~Z. Kolter, ``Tofu: A task of fictitious unlearning for llms,'' \emph{arXiv preprint arXiv:2401.06121}, 2024.

\bibitem{eldan2023s}
R.~Eldan and M.~Russinovich, ``Who's harry potter? approximate unlearning in llms,'' \emph{arXiv preprint arXiv:2310.02238}, 2023.

\bibitem{luo2023bc4llm}
H.~Luo, J.~Luo, and A.~V. Vasilakos, ``Bc4llm: Trusted artificial intelligence when blockchain meets large language models,'' \emph{arXiv preprint arXiv:2310.06278}, 2023.

\bibitem{gong2023dynamic}
Y.~Gong, ``Dynamic large language models on blockchains,'' \emph{arXiv preprint arXiv:2307.10549}, 2023.

\bibitem{lin2024blockchain}
Y.~Lin, Z.~Gao, H.~Du, J.~Ren, Z.~Xie, and D.~Niyato, ``Blockchain-enabled trustworthy federated unlearning,'' \emph{arXiv preprint arXiv:2401.15917}, 2024.

\bibitem{mboma2023assessing}
J.~G.~M. Mboma, O.~T. Tshipata, W.~V. Kambale, and K.~Kyamakya, ``Assessing how large language models can be integrated with or used for blockchain technology: Overview and illustrative case study,'' in \emph{2023 27th International Conference on Circuits, Systems, Communications and Computers (CSCC)}.\hskip 1em plus 0.5em minus 0.4em\relax IEEE, 2023, pp. 59--70.

\bibitem{malhotra2024blockchain}
D.~Malhotra, P.~Saini, and A.~K. Singh, ``Blockchain-based proof-of-authenticity frameworks for explainable ai,'' \emph{Multimedia Tools and Applications}, vol.~83, no.~13, pp. 37\,889--37\,911, 2024.

\bibitem{zhang2023privacyeafl}
M.~Zhang, S.~Chen, J.~Shen, and W.~Susilo, ``Privacyeafl: Privacy-enhanced aggregation for federated learning in mobile crowdsensing,'' \emph{IEEE Transactions on Information Forensics and Security}, 2023.

\bibitem{zong2023relac}
J.~Zong, C.~Wang, J.~Shen, C.~Su, and W.~Wang, ``Relac: Revocable and lightweight access control with blockchain for smart consumer electronics,'' \emph{IEEE Transactions on Consumer Electronics}, vol.~70, no.~1, pp. 3994--4004, 2023.

\bibitem{li2023blockchain}
Y.~Li, J.~Shen, S.~Ji, and Y.-H. Lai, ``Blockchain-based data integrity verification scheme in aiot cloud--edge computing environment,'' \emph{IEEE Transactions on Engineering Management}, 2023.

\bibitem{wooldridge2009introduction}
M.~Wooldridge, \emph{An introduction to multiagent systems}.\hskip 1em plus 0.5em minus 0.4em\relax John Wiley \& Sons, 2009.

\bibitem{watkins1992q}
C.~J. Watkins and P.~Dayan, ``Q-learning,'' \emph{Machine learning}, vol.~8, no. 3-4, pp. 279--292, 1992.

\bibitem{yin2020fedloc}
F.~Yin, Z.~Lin, Q.~Kong, Y.~Xu, D.~Li, S.~Theodoridis, and S.~R. Cui, ``Fedloc: Federated learning framework for data-driven cooperative localization and location data processing,'' \emph{IEEE Open Journal of Signal Processing}, vol.~1, pp. 187--215, 2020.

\bibitem{politou2019blockchain}
E.~Politou, F.~Casino, E.~Alepis, and C.~Patsakis, ``Blockchain mutability: Challenges and proposed solutions,'' \emph{IEEE Transactions on Emerging Topics in Computing}, vol.~9, no.~4, pp. 1972--1986, 2019.

\bibitem{akbar2021distributed}
N.~A. Akbar, A.~Muneer, N.~ElHakim, and S.~M. Fati, ``Distributed hybrid double-spending attack prevention mechanism for proof-of-work and proof-of-stake blockchain consensuses,'' \emph{Future Internet}, vol.~13, no.~11, p. 285, 2021.

\bibitem{wan2022zk}
Z.~Wan, Y.~Zhou, and K.~Ren, ``zk-authfeed: Protecting data feed to smart contracts with authenticated zero knowledge proof,'' \emph{IEEE Transactions on Dependable and Secure Computing}, vol.~20, no.~2, pp. 1335--1347, 2022.

\end{thebibliography}

\end{document}